\documentclass[12pt]{article}

\usepackage[dvips,letterpaper,body={16.5cm,22.9cm},noheadfoot]{geometry}
\usepackage{graphicx}
\newcommand{\myskip}{\vspace{\baselineskip}}
\newcommand{\mysection}[1]{\par\myskip\noindent\textbf{#1}\myskip\par}

\begin{document}

\begin{center}
\large {\bf Kondo effect in a two-level quantum dot coupled to an
external fermionic reservoir}
\bigskip

\normalsize
A. L. Chudnovskiy and S. E. Ulloa \\
Department of Physics and Astronomy and Condensed Matter and
Surface Sciences, Ohio University, Athens, OH 45701-2979 USA

\bigskip


{\bf Abstract}\\
\bigskip
\parbox{4.5in}{We investigate theoretically the linear
conductance of a two-level quantum dot as a function of the gate
voltage and different strength of coupling to the external
electronic system (the reservoir). Apart from the weak coupling
regime, characterized by the Kondo-enhancement of the conductance
in the spinful ground state, a strong coupling regime, which can
be called {\em mixed-valence} (MV), is found. This regime is
characterized by a qualitative change of the energy level
structure in the dot, resulting in {\it sub}- and {\it
super-tunneling} coupling of the levels, which in turn yield a
novel temperature dependence of the Kondo effect in the quantum
dot.}

\end{center}

The Kondo effect arises from the coherent screening between a
localized spin and that of surrounding mobile electrons, producing
for example anomalous transport properties in metals with magnetic
impurities \cite{Hewson}.  Recently, however, there has been a
great deal of experimental activity in systems where {\em an
individual} localized spin is probed directly in quantum dots
defined in semiconductor systems \cite{DavidGG,SC}.

Previous investigations on both single- and multi-level models
have uncovered interesting features of the Kondo effect, including
Kondo peaks in the conductance and the associated density of
states, their temperature dependence, and other features
\cite{DavidGG,Pohjola}.   However, the typical approximation made
in models of multilevel quantum dots with discrete energy levels
is to neglect the strong mixing between the energy levels of the
dot due to the interaction with the external fermionic system.
This approach incorporates the external fermionic system only as a
broadening of the levels in the quantum dot \cite{Pohjola}.

In this paper we analyze the Kondo effect in a quantum dot coupled
to an external fermionic system in addition to the coupling to the
measuring leads. We explicitly take into account the mixing of the
states in the dot due to the coupling to an external reservoir. A
typical experimentally accessible example of such a system is a
dot coupled to {\em three} leads with one lead playing the role of
the reservoir and drawing no current. It is assumed that the
coupling to the fermionic reservoir can be varied independently of
the coupling to the measuring leads. In the case of weak coupling,
the reservoir just leads to broadening of the energy levels in the
dot, whereas strong coupling results in a qualitative
rearrangement of the energy levels which influences the Kondo
effect in a nontrivial manner.

In what follows we consider a two-level quantum dot in the linear
(zero bias) regime at zero temperature. The Hamiltonian of the
model can be written as
\begin{equation}
H = \sum_{l,\sigma} \{E_l \hat{c}^+_{l,\sigma}\hat{c}_{l,\sigma}
+U_1\sum_{l\neq l'}\hat{n}_l\hat{n}_{l'}
+U\hat{n}_{l\uparrow}\hat{n}_{l\downarrow}\}
+\sum_{\nu=1}^3\gamma_\nu\sum_{l,\sigma}
(\hat{c}^+_{l,\sigma}\hat{a}_{\nu,0,\sigma}+
\hat{a}^+_{\nu,0,\sigma}\hat{c}_{l,\sigma}) +
\sum_{\nu=1}^3 H^\nu_F(\hat{a}^+_{\nu,r,\sigma}, \hat{a}_{\nu,r,\sigma}).
\label{H}
\end{equation}
Here, the fermionic operators $\hat{c}^+_{l,\sigma},
\hat{c}_{l,\sigma}$ describe the states in the dot with orbital
index $l=1,2$, and spin index $\sigma$.
$\hat{n}_{l,\sigma}=\hat{c}^+_{l,\sigma}\hat{c}_{l,\sigma}$ is the
particle number operator of the state $(l,\sigma)$, and
$\hat{n}_l=\hat{n}_{l\uparrow}+\hat{n}_{l\downarrow}$. The dot is
coupled by tunnel couplings $\gamma_\nu$ to the two {\em
measuring} leads, the right ($R$, $\nu=1$) and the left ($L$,
$\nu=2$). The third lead couples the dot to a fermionic reservoir
via the coupling $\gamma_3$. In what follows we consider the
symmetric case, $\gamma_1=\gamma_2$, $\gamma_3=\gamma$.
$H^\nu_F$'s denote the Hamiltonians of the leads. The fermions at
the space position $r=0$ are considered to be coupled directly to
the dot.

The basic interactions in the quantum dot are the density-density
repulsion between charges on {\em different} energy levels via the
interaction constant $U_1$, and the Hubbard-like repulsion $U$
between electrons on a given level of the dot.  The interaction
constants depend mostly on overlap integrals between the wave
functions of the different energy levels. Therefore, they can be
changed by changing the shape of the quantum dot, which is
experimentally accessible \cite{DavidGG}.

To describe the effect of the quantum dot on the transport through
the measuring leads, we integrate out the degrees of freedom
related to the dot and reservoir.  We then obtain the effective
Lagrangian for the measuring leads only, which is given by
\begin{equation}
L_{eff}=\sum_{\nu=R,L}L^0_\nu(\bar{a}^+_{\nu,r,\sigma},
a_{\nu,r,\sigma})-\gamma_1^2\sum_{\omega,\omega'}\sum_{l,l'}
\left(\sum_{\nu=R,L} \bar{a}^{\omega +}_{\nu,r=0,\sigma}\right)
\Upsilon^{\omega\omega'}_{ll',\sigma\sigma'}
\left(\sum_{\nu'=R,L}a^{\omega' +}_{\nu',r=0,\sigma'}
\right),
\label{Leff1}
\end{equation}
where $\Upsilon_{ll'}(i\omega_n)\equiv \langle\hat{c}_l(i\omega_n)
\hat{c}_{l'}^+(i\omega_n)\rangle$ is the fermionic one-particle
Green's function of the dot coupled to the reservoir but {\it
isolated} from the measuring leads.
$L^0_\nu(\bar{a}^+_{\nu,r,\sigma}, a_{\nu,r,\sigma})$ denotes the
Lagrangian of the isolated lead ($\nu=R,L$), and $\omega,\omega'$
are Matsubara frequencies. To find $\Upsilon_{ll'}(i\omega_n)$, 
we integrate out the reservoir in the Hamiltonian (\ref{H}). 
In the result,  the effective
Lagrangian acquires a new term $i\kappa
\sum_{l,l'}\sum_{\sigma=\uparrow,\downarrow}
\bar{c}_{l\sigma}^\omega c_{l'\sigma}^\omega$ that describes 
broadening and mixing of the energy levels in the quantum dot. 
For Matsubara frequencies much less than
the bandwidth of the reservoir $D$, we obtain $\kappa \approx \
{\rm sign} (\omega_n)\pi\gamma^2\rho_0$, where $\rho_0$ is the
density of states in the reservoir \cite{Hewson,AC-SU}.

We treat the many-body interaction in the dot in a Hartree
approximation (the spin is assumed frozen in the $z$-direction)
with subsequent expansion in spin fluctuations to take into
account the Kondo effect. In the mean field (Hartree)
approximation, the two level quantum dot with interactions is
replaced by a {\em non-interacting} quantum dot with quasi-energy
levels that depend on the ground state of the quantum dot. This
approach allows one to describe correctly the properties of the
quantum dot in the ground state and its gapless excitations,
including the Kondo effect, which is the main aim of this
treatment. Due to the mixing and broadening of the energy levels
in the dot coupled to the reservoir, the effective energy levels
for each spin projection become $z^\sigma_\pm =
\left(\epsilon_{1\sigma} +\epsilon_{2\sigma}\pm
\sqrt{(\epsilon_{1\sigma}-\epsilon_{2\sigma})^2-2\kappa^2}\right)/2
-i\kappa \,{\rm sign} (\omega_n), $ where $\epsilon_{l\sigma}$
denote the effective energy level for the isolated dot. Detailed
discussion of the mean-field equations and their solutions is
given in Ref. \cite{AC-SU}.

In the limit of zero level mixing (zero tunnelling $\gamma$), the
values of $z_\pm$ coincide with $\epsilon_{1,2}$.  Small level
mixing $\kappa$ leads to small deviations from these solutions. On
the other hand, in the large level mixing regime, when the
condition $2\kappa > |\epsilon_1-\epsilon_2|$ is satisfied, the
quasi-level structure changes qualitatively. The quasienergies are
given by $
z_{1,2}=(\epsilon_1+\epsilon_2)/2-i\kappa\left(1\mp\sqrt{1-
(\epsilon_1-\epsilon_2)^2/(4\kappa^2)}\right) {\rm sign}
(\omega_n)$, corresponding to two degenerate levels:  one strongly
broadened ($z_2$) ``supertunneling'' level, and the other one a
``subtunneling'' level with strongly suppressed broadening ($z_1$)
\cite{Shab-Ulloa}. The qualitative change of the effective energy
level structure by going from small to large coupling to the
external reservoir has a profound influence on the transport
properties of the quantum dot, particularly in the Kondo ground
state.

The physics of the Kondo effect is taken into account by
considering spin fluctuations around the spinful saddle point. 
The charge fluctuations are massive and can be omitted \cite{AC-SU}.
The only modes of the spin fluctuation field $Q_{ll'}$  that can
become massless and give rise to the Kondo effect are described by
the correlators $ \langle(\sigma^+\otimes Q_{ll})(\sigma^-\otimes
Q_{ll}) \rangle_{\omega}= \langle(\sigma^-\otimes
Q_{ll})(\sigma^+\otimes Q_{ll})
\rangle_{-\omega}=(16U^2)/[\omega+4U(1-\phi_{ll}/\pi)] $ with
$\phi_{ll}=\arccos(-z'_{l\downarrow}/|z_{l\downarrow}|)-
\arccos(-z'_{l\uparrow}/|z_{l\uparrow}|)$. In the case of small
broadening $\kappa$, $|z_{l\sigma}|\approx |z'_{l\sigma}|$. If the
state of the level $l$ is spinful, then $z'_{l\uparrow}<0$ and
$z'_{l\downarrow}>0$, hence $\phi_{ll}=\pi$ and the correlator
becomes massless. In the strong-coupling regime, which we will
call ``mixed valence'' (MV), the subtunneling level has a very
small broadening, and hence the fluctuations of the spin field for
this level are massless, whereas the supertunneling level is
strongly broadened and its spin fluctuations are massive.
Physically, the electron on the supertunneling level is
effectively {\em delocalized}, thus giving no Kondo effect in the
usual sense, but having a different contribution in that regime
(see below).

Expanding the function $\Upsilon$ in the Lagrangian
(\ref{Leff1}) to lowest order of the fluctuation matrix $\hat{Q}$,
we obtain the Kondo part of the interaction, from which we
identify the Kondo constants. In the Kondo regime, $\kappa\ll
|\epsilon_{1\sigma}-\epsilon_{2\sigma}|$, the Kondo coupling for
the level $l$ assumes the form $ J_K^{l} =
-{4U\gamma_1^2}/{\epsilon_{l\uparrow} \epsilon_{l\downarrow}} \,
$. Since $\epsilon_{l\uparrow}\epsilon_{l\downarrow}<0$, the Kondo
coupling is antiferromagnetic, $J_K^{l}>0$. In the MV regime, only
the subtunneling level contributes to the Kondo effect. For this
level (let it be $l=1$), we have $ J_K^{sub}\approx -
\frac{4U\gamma_1^2(E_1-E_2)^2} {2\kappa^2
(\epsilon_{1\uparrow}+\epsilon_{2\uparrow})(\epsilon_{1\downarrow}+
\epsilon_{2\downarrow})} \, $. The Kondo coupling in the MV regime
becomes ferromagnetic, and it is weaker that in the Kondo regime
by the factor $(E_1-E_2)^2/(8\kappa^2)$. The Kondo temperature can
be evaluated as $T_K\sim D \exp[-1/(\nu_F J_K)$, where $\nu_F$ is
the density of states at the Fermi level in the leads. It follows 
from the expressions for the Kondo couplings  that whereas in the
Kondo regime $T_K$ is determined by the position of a given level,
in the MV regime $T_K$ is determined by the level mixing $\kappa$,
as the positions of both levels enter symmetrically.

\begin{figure}
\includegraphics[width =17cm]{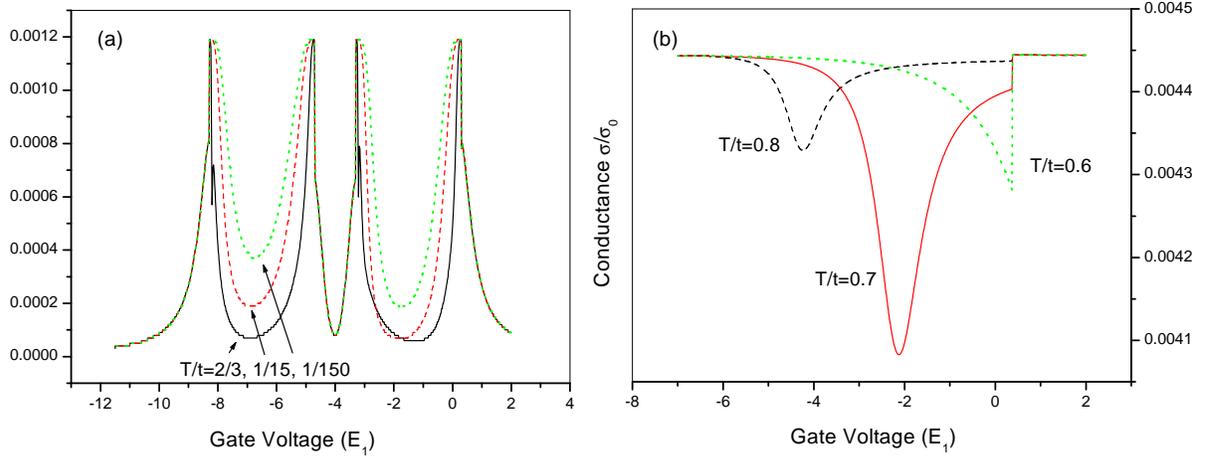}
\caption{Conductance of the quantum dot versus the gate voltage
($E_1$). $U=3, U_1=1$ at three different temperatures. (a) The
Kondo (weak coupling) regime: $t=15, \gamma_1=\gamma=0.5$. (b) The
mixed valence (strong coupling) regime: $t=50,
\gamma_1=\gamma=4$.} \label{condKo}
\end{figure}

The behavior of the conductance versus gate voltage in the Kondo
regime at different temperatures is illustrated in Fig.\
\ref{condKo}(a). The conductance is normalized by the value of the
conductance through the leads. In the spinful state, the
nonresonant conductance is enhanced due to the Kondo effect. 
The conductance drops with temperature, which is also
characteristic for the Kondo effect temperature dependence. The
levels in the dot are widely separated, so that one observes two
subsequent Kondo states as the gate voltage is varied. The
conductance versus the gate voltage in the MV regime is shown in
Fig.\ \ref{condKo}(b).  In this case, the enhanced scattering due
to the Kondo effect involves scattering processes {\em into} the
external reservoir, which leads to a {\it reduction} of the
conductance through the dot. Therefore, in the spinful state we
have a Kondo ``dip'' of the conductance instead of a Kondo peak.
The Kondo dip in the conductance has an  unusual temperature
dependence.  Notice that, in contrast to the Kondo regime, the
position of the Kondo dip {\em shifts} with temperature in the MV
regime. As it has been discussed above, in the MV regime, the
Kondo contribution to the conductance is generated only by the
subtunneling level. However, because of the large background
conductance by the supertunneling level (through which the dot is
practically open), the experimental identification of the Kondo
dips shown in Fig.\ \ref{condKo}(b) might be difficult, especially
in comparison with impurity effects in these systems (which would
yield universal conductance fluctuations, for example).  However,
the large shifts in the dip position with temperature may help a
great deal to isolate this effect.

We emphasize that the predicted effects would not be present for a
single-level dot, where the broadening of the level suppresses the
Kondo effect for strong coupling. Therefore, the mixing of the
different energy levels in the dot by the coupling to the external
fermionic system is essential for the observation of the Kondo
effect in the MV regime. \\

We acknowledge support from the US Department of Energy grant no.\
DE--FG02--87ER45334.

\mysection{References} 

\end{document}